\documentclass[%
 aip,
 amsmath,amssymb,
 reprint,%
]{revtex4-1}

\usepackage{graphicx}
\usepackage{dcolumn}
\usepackage{bm}

\usepackage[utf8]{inputenc}
\usepackage[T1]{fontenc}
\usepackage{mathptmx}
\usepackage{etoolbox}
\usepackage{bigints}
\usepackage[dvipsnames]{xcolor}
\usepackage{multirow}
\usepackage{graphicx}
\usepackage[export]{adjustbox}

\makeatletter
\def\@email#1#2{%
 \endgroup
 \patchcmd{\titleblock@produce}
  {\frontmatter@RRAPformat}
  {\frontmatter@RRAPformat{\produce@RRAP{*#1\href{mailto:#2}{#2}}}\frontmatter@RRAPformat}
  {}{}
}%
\makeatother
\begin{document}

\preprint{AIP/123-QED}

\title[Ethane hydrate dissociation line from molecular simulation]{Prediction of the three-phase coexistence line of the ethane hydrate from molecular simulation}

\author{Paula Gómez-Álvarez}
\affiliation{Laboratorio de Simulaci\'on Molecular y Qu\'imica Computacional, CIQSO-Centro de Investigaci\'on en Qu\'imica Sostenible and Departamento de Ciencias Integradas, Universidad de Huelva, 21006 Huelva Spain}

\author{Miguel J. Torrej\'on}
\affiliation{Laboratorio de Simulaci\'on Molecular y Qu\'imica Computacional, CIQSO-Centro de Investigaci\'on en Qu\'imica Sostenible and Departamento de Ciencias Integradas, Universidad de Huelva, 21006 Huelva Spain}

\author{Jes\'us Algaba}
\affiliation{Laboratorio de Simulaci\'on Molecular y Qu\'imica Computacional, CIQSO-Centro de Investigaci\'on en Qu\'imica Sostenible and Departamento de Ciencias Integradas, Universidad de Huelva, 21006 Huelva Spain}

\author{Felipe J. Blas$^*$}
\affiliation{Laboratorio de Simulaci\'on Molecular y Qu\'imica Computacional, CIQSO-Centro de Investigaci\'on en Qu\'imica Sostenible and Departamento de Ciencias Integradas, Universidad de Huelva, 21006 Huelva Spain}

\begin{abstract}

We investigate the three-phase coexistence line of ethane (C$_2$H$_6$) hydrate through molecular dynamics simulations using the direct coexistence approach. In this framework, C$_2$H$_6$ sI hydrate, aqueous, and pure guest phases are constructed within a single simulation box, allowing us to monitor their mutual stability. From the temporal evolution of the potential energy, we identify the equilibrium temperature ($T_3$) at which all three phases coexist, across pressures ranging from $1000$ to $4000\operatorname{bar}$, in accordance with available experimental data. Simulations are performed with the GROMACS package (version 2016, double precision) in the $NPT$ ensemble. Water and C$_2$H$_6$ molecules are represented using the TIP4P/Ice and TraPPE-UA models, respectively, while unlike non-bonded interactions are computed with the Lorentz–Berthelot combining rule. Dispersive Lennard-Jones and Coulomb interactions are truncated at $1.6\operatorname{nm}$, with long-range Coulombic contributions treated via Particle-Mesh Ewald summation. The predicted three-phase coexistence line shows excellent agreement with experimental measurements within the investigated pressure range. These results demonstrate the suitability of the direct coexistence methodology, combined with established molecular models, for reproducing hydrate dissociation behavior in systems that have received little prior computational attention.
\end{abstract}

\maketitle
$^*$Corresponding author: felipe@uhu.es

%

\section{Introduction}

In the late eighteenth century, several natural philosophers noted the unexpected formation of ice-like solids—even at temperatures above the freezing point of water—when certain gases were bubbled through cold water or when such mixtures were frozen. Sir Humphry Davy was the first scientist to identify these substances as compounds of water and gas, introducing the term ``gas hydrates''. Following more than a century of continued investigation, these materials were ultimately recognized as clathrates: crystalline frameworks in which small guest molecules are encapsulated within hydrogen-bonded water cages, as exemplified by methane (CH$_{4}$) and carbon dioxide (CO$_{2}$) under appropriate thermodynamic conditions.~\cite{Sloan2008a,Barnes2013a,Ripmeester2022a}

Clathrate frameworks also occur in intermetallic compounds,~\cite{Dolyniuk2016a,Kasper1965a} where metallic guest ions are enclosed by group 14 (Si, Ge, Sn) frameworks. More recently, clathrate-like assemblies have been realized from colloidal particles,~\cite{Lin2017a,Lee2019a} enabling larger cages and mesoscale phenomena relevant to biological and photonic systems.

In both gas hydrates and more complex clathrates, host–guest interactions govern stability and functionality, underpinning their utility in natural gas capture~\cite{Sloan2008a} and storage.~\cite{Mao2002a} CH$_{4}$ hydrates have drawn particular attention as a potential energy resource due to their extensive natural reserves,~\cite{Sloan2008a,Ripmeester2022a,Sloan2003a,Koh2012a,Chong2016a,Zheng2020a,Bourry2007a,Yousuf2004a,Makogon1997,Aghajari2019a} and as a significant climate factor, acting as a large greenhouse gas reservoir susceptible to destabilization under environmental change.~\cite{Ripmeester2022a,Sloan2008a,Manakov2003a,Makino2005a,Kvenvolden1988a,Koh2012a,Sloan2003a} Additional interest stems from their potential in CO$_{2}$ sequestration~\cite{Ohgaki1996a,Yang2014a,Ricaurte2014a} and in gas storage~\cite{Kvamme2007a} and transport technologies.~\cite{Chihaia2005a,Peters2008a,English2009a}

Accurate knowledge of the thermodynamic stability of gas hydrates is essential for their effective deployment in energy, sequestration, gas storage, and transportation applications. Over recent decades, numerous studies have experimentally determined the hydrate dissociation line, which typically corresponds to a three-phase equilibrium in a two-component system, where hydrate, aqueous, and guest-rich gas or liquid phases coexist—depending on the specific guest molecule involved. Comprehensive discussions can be found in the authoritative monograph by Sloan and Koh,~\cite{Sloan2008a} as well as in the recent volume by Ripmeester and Alavi,~\cite{Ripmeester2022a} which offers a detailed review of hydrate phase behavior. From a structural point of view, Matsumoto and Tanaka realized that the arrangement of the guest molecules in the sI and sII hydrate structures is identical to the A15 and C15, respectively,  atomic arrangement in the Frank-Kasper structure classification.~\cite{Frank1958a,Matsumoto2011a,Matsumoto2022a} Following this approach, clathrate hydrates can also be viewed from a geometrical perspective, and can be classified as Frank–Kasper structures, which are space-filling packings of “nearly equal” spheres with tetrahedrally close-packed (TCP) topology. This “nearly equal spheres” approach offers a unifying geometrical principle that rationalizes the diversity of hydrate frameworks and provides a natural link between their topological characteristics and the stability trends governed by guest size and occupancy.~\cite{Matsumoto2011a,Matsumoto2022a,Muromachi2024a,Chen2023a}

Molecular simulation is an alternative method to determine the dissociation lines of hydrates.~\cite{Allen2017a,Frenkel2002a,Sloan2008a,Ripmeester2022a} In particular, the direct coexistence method (DC) proposed by Ladd and Woodcock~\cite{Ladd1977a,Ladd1978a} was successfully extended to the case of hydrates by Carlos Vega and co-workers,~\cite{Conde2010a,Conde2013a} who applied it to CH$_{4}$.
Thanks to the brilliant and pioneering work of Conde and Vega,~\cite{Conde2010a} this approach gave rise to a large number of studies in which the dissociation line of different hydrates was obtained using the direct coexistence method.~\cite{Conde2010a,Jensen2010a,Miguez2015a,Michalis2015a,Constandy2015a,Perez-Rodriguez2017a,Fernandez-Fernandez2019a,Fernandez-Fernandez2021a,Blazquez2023b,Algaba2024a,Algaba2024b,Borrero2025a,Torrejon2025a} Recently, Tanaka and collaborators~\cite{Tanaka2018a} have proposed an alternative approach, the solubility method, which provides a different route to determine the dissociation lines of hydrates. Several authors have employed this method to predict dissociation lines.~\cite{Grabowska2022a,Grabowska2022b,Algaba2023a,Algaba2023b,Torrejon2024b,Torrejon2024b} Hydrate properties including growth, cage occupancy, order parameters, interfacial free energies, and nucleation have also been addressed through molecular simulation in prior studies.~\cite{Alavi2005a,Alavi2006a,Alavi2007a,Grabowska2022b,Zeron2025a,Algaba2022b,Zeron2022a,Romero-Guzman2023a,Torrejon2024a,Zeron2024c,Garcia2025a,DeFever2017a,Brumby2016a,Brumby2019a,Krishnan2022a,Bagherzadeh2015a,Katsumasa2007a,Cao2020a,Hakim2010a,Nakayama2009a,Fang2017a,Fang2023a,Fang2024a,Matsuo2012a,Yagasaki2014a,Walsh2011a,Knott2012a,Barnes2013a,Barnes2014a,Barnes2014b,Jacobson2010a,Jacobson2010b,Jacobson2010c,Jacobson2010d,Jacobson2011a,Jiao2021a,Hao2023a,Qiu2018a,Wang2023a,Liang2011a,Liang2013a,Waage2017a,Tung2011a}

Although hydrates have diverse applications, including gas separation, transport, and storage, natural gas hydrates attract particular attention due to their significance as an energy resource. The main components of natural gas are CH$_{4}$, CO$_{2}$, and light linear alkanes such as ethane (C$_{2}$H$_{6}$). Interestingly, although the dissociation line of C$_{2}$H$_{6}$ hydrate has been obtained experimentally by several authors,~\cite{Roberts1940a,Deaton1946a,Reamer1952a,Galloway1970a,Falabella1974a,Holder1980a,Holder1982a,Ng1985a,Avlonitis1988a,Song1989a,Nakano1998b,Yang2000a,Morita2000a} to the best of our knowledge, it has not yet been investigated from a molecular perspective.

Gas hydrates generally crystallize into one of three well-characterized structures: sI, sII, or sH. Small guest molecules such as CH$_{4}$ and CO$_{2}$ typically form sI hydrates while others, including nitrogen (N$_{2}$), hydrogen (H$_{2}$), and larger species like tetrahydrofuran (THF), more commonly form sII hydrates.~\cite{Sloan2008a,Ripmeester2022a}

The stable structure of C$_{2}$H$_{6}$ hydrate, as in the case of CH$_{4}$ and CO$_{2}$ hydrates, is structure sI. The sI unit cell consists of $46$ water molecules arranged into two pentagonal dodecahedral (D or $5^{12}$) cages and six tetrakaidecahedral (T or $5^{12}6^{4}$) cages. Under the assumption of single full occupancy, $8$ C$_{2}$H$_{6}$ molecules would occupy all the cages in each unit cell. Until the turn of the twenty-first century, it was generally believed that C$_{2}$H$_{6}$ molecules occupied only the large T cages, yielding a composition of $6$ C$_{2}$H$_{6}$ molecules per $46$ water molecules, i.e., one C$_{2}$H$_{6}$ molecule per $7.67\approx46/6$ water molecules. However, Morita and co-workers demonstrated that molecules can occupy both types of cages.~\cite{Morita2000a} In particular, it had long been assumed that C$_{2}$H$_{6}$ was too large to fit into the smaller D cages due to the unfavorable ratio of molecular diameter to cage diameter.~\cite{Sloan2008a} C$_{2}$H$_{6}$ hydrate was therefore regarded as a typical clathrate hydrate consisting of filled T cages and empty D cages. Using Raman spectroscopy of the C--C stretching vibration of C$_{2}$H$_{6}$, Morita et al.~\cite{Morita2000a} showed that C$_{2}$H$_{6}$ is in fact encapsulated in both T and D cages. The Raman signal corresponding to C--C vibrations in D cages is weak at pressures below $100\ \mathrm{MPa}$, but becomes significant at higher pressures. See Fig.~4 of the work of Morita and collaborators for further details.~\cite{Morita2000a} This provided the first direct spectroscopic evidence of D-cage occupancy by C$_2$H$_6$. Following Morita and co-workers, we assume in this work full occupancy in the sI structure of the C$_{2}$H$_{6}$ hydrate ($8$ C$_{2}$H$_{6}$ molecules per $46$ water molecules).

The objective of this work is to determine the dissociation line of the hydrate–water–liquid C$_{2}$H$_{6}$ system over a wide pressure range, using the direct coexistence technique first brilliantly implemented by Carlos Vega and co-workers.~\cite{Conde2010a,Conde2013a} C$_{2}$H$_{6}$ hydrate, like some other hydrates, exhibits two dissociation lines: a hydrate–water–liquid ethane line at high pressures and a hydrate–water–vapor ethane line at low pressures. In this study, we focus exclusively on the stable high-pressure branch. Simple but effective models are employed that capture the essential features required for a reliable prediction of the dissociation line: a water model that successfully reproduces the melting point of water at ambient pressure, and a robust ethane model. Specifically, we use the TIP4P/Ice model for water~\cite{Abascal2005b} and the TraPPE-UA model for C$_{2}$H$_{6}$.~\cite{Martin1998a,Shah2017a}

The remainder of this paper is organized as follows. Section~II presents the molecular models, methodology, and simulation details. The results and their discussion are given in Section~III, and the conclusions are summarized in Section IV.

\begin{table}
\caption{Non-bonded interaction parameters and geometry details of TIP4P/Ice water~\cite{Abascal2005b} and TraPPE-UA C$_2$H$_6$\cite{Martin1998a,Shah2017a}
 molecular models employed in this work.}
\centering
\begin{tabular}{lccccc}
\hline\hline
Atom &$\sigma(\text{\AA})$ & $\varepsilon/k_B(\text{K})$ & $q$(e) & \multicolumn{2}{c}{Geometry details} \\
\hline
\multicolumn{6}{c}{Water (TIP4P/Ice)} \\
\hline
O            & 3.1668 & 106.1 & - & $d_\text{OH}$ (\AA) & 0.9572\\
H & - & - & 0.5897 & H--O--H (º) & 104.5\\
M  & - & - & -1.1794 & $d_\text{OM}$ (\AA) & 0.1577\\
\hline
\multicolumn{6}{c}{C$_2$H$_6$ (TraPPE-UA)} \\
\hline
CH$_3$ & 3.75 & 98.0 & - & $d_{\text{CH}_3-\text{CH}_3}$ (\AA) & 1.54 \\
\hline
\hline
\end{tabular}
\label{Table-model}
\end{table}

\section{Molecular models, methodology, and simulation details}

Molecular dynamics simulations are carried out using the GROMACS software package (version 2016, double precision). Water and ethane molecules are modeled using the well-known TIP4P/Ice~\cite{Abascal2005b} and TraPPE-UA~\cite{Martin1998a,Shah2017a} molecular models. In all cases, the non-bonded unlike interactions are obtained through the Lorentz-Berthelot combination rule, and no extra cross-interaction modification is applied. A summary of the most relevant molecular model details is presented in Table \ref{Table-model}.

\begin{figure*}
\includegraphics[width=\textwidth]{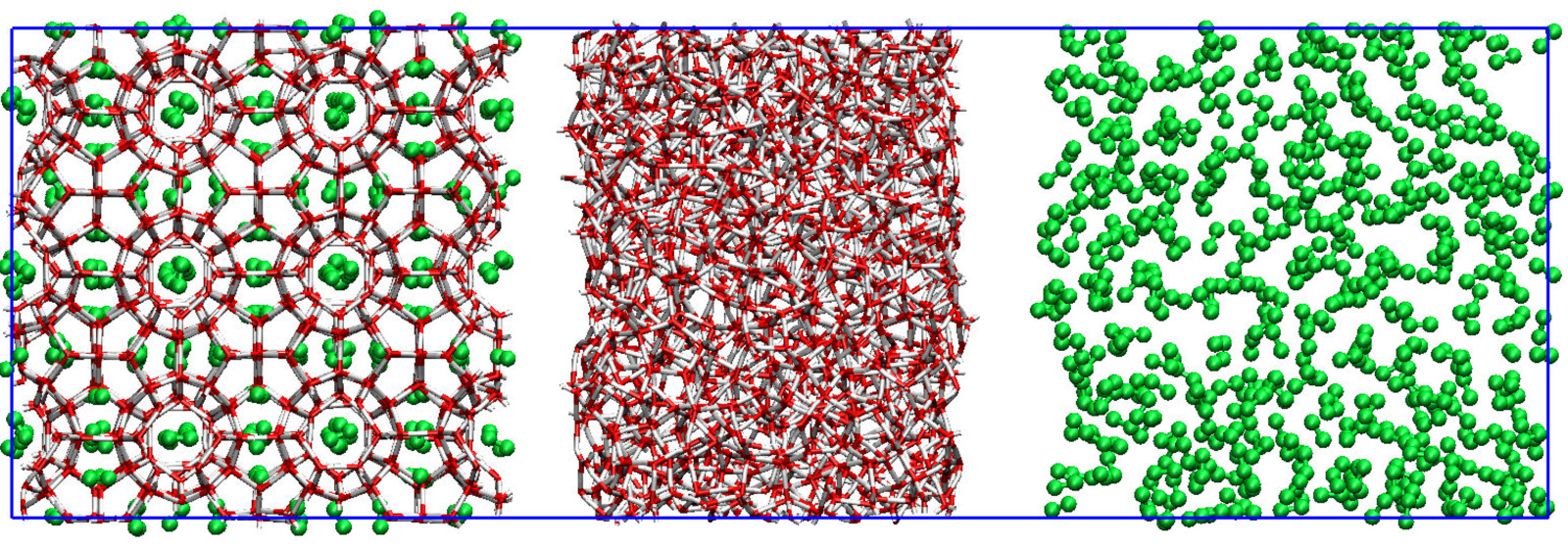}   
\caption{Representation of the initial simulation box used in this work. green spheres represent the C$_2$H$_6$ molecules and red and white licorice representations correspond to water molecules. From left to right, the simulation box is conformed by a hydrate phase, a water phase, and a C$_2$H$_6$ phase.}
\label{snapshot}
\end{figure*}

The dispersive Lennard-Jones (LJ) and Coulomb interactions are truncated through a cutoff value of $1.6\,\text{nm}$. Long-range corrections for the dispersive LJ interactions are not applied, but particle-mesh Ewald (PME)~\cite{Essmann1995a} corrections are used for the Coulombic potential. Notice that the choice of using a long cutoff value without long-range corrections is far from being arbitrary. As some of the authors of this work have claimed in a previous study, using a small cutoff value could lead to an incorrect prediction of the three-phase dissociation temperature.~\cite{Algaba2024b} Although the combination of a small cutoff value and inhomogeneous PME long-range corrections for the LJ dispersive interactions provides a correct estimation of the $T_3$ value at the same time that reduces the computational cost of the simulations, in this work we decide to use a large cutoff without any long-range correction for the LJ dispersive interactions. The use of a large cutoff value provides the same results as those obtained when long-range corrections are applied.~\cite{Algaba2024b}Although the computational cost of the simulations is higher when large cutoff values are used instead of long-range corrections, in GROMACS it is not possible to apply long-range corrections for the LJ dispersive interactions at the same time that the Lorentz-Berthelot combination rule is modified to improve the agreement between experiments and simulations.~\cite{Miguez2015a,Constandy2015a,Algaba2023a,Algaba2024a,Algaba2024b,Algaba2024c,Michalis2022a,Algaba2023b,Torrejon2024b,Waage2017a} For this reason, we have decided to use a large $1.6\,\text{nm}$ cutoff value without LJ corrections just in case the Lorentz-Berthelot combination rule must be modified to improve the $T_3$ predictions.

The $T_3$ or dissociation temperature of the C$_2$H$_6$ sI hydrate is determined from $1000$ to $4000\,\text{bar}$ using the direct coexistence technique.~\cite{Ladd1977a,Ladd1978a,Miguez2015a,Costandy2015a,Michalis2015a,Blazquez2024a,Algaba2024a,Algaba2024b} Following this methodology, an initial C$_2$H$_6$ sI hydrate, aqueous, and pure guest phases are assembled in the same simulation box, with a central phase surrounded on each side by the other two phases, which allows three-phase coexistence due to periodic boundary conditions. By performing $NPT$ simulations, it is possible to determine the temperature at which the three phases coexist. At a certain pressure value, if the temperature fixed during the simulation is higher than the $T_3$ ($T>T_3$), then the hydrate phase melts since it becomes unstable. On the contrary, if the temperature fixed during the simulation is lower than the $T_3$ ($T<T_3$), then the hydrate phase grows since the aqueous phase becomes unstable. The $T_3$ value is determined as the intermediate temperature between the highest $T$ value at which the hydrate phase grows and the lowest $T$ value at which the hydrate phase melts. Uncertainties are estimated by subtracting these two temperatures and dividing by two.~\cite{Conde2010a,Miguez2015a}

To construct the hydrate unit cell, we used the crystallographic parameters reported by Yousuf \emph{et al}.~\cite{Yousuf2004a} Additionally, the water molecules in C$_{2}$H$_{6}$ hydrate exhibit proton disorder.~\cite{Sloan2008a}  To account for this, we generated solid configurations of the sI hydrate using the algorithm developed by Buch \emph{et al}.,~\cite{Buch1998a} which enforces the Bernal–Fowler rules~\cite{Bernal1933a} and ensures a net dipole moment that is zero or nearly zero. The initial hydrate phase is then constructed by replicating the C$_2$H$_6$ sI hydrate unit cell 3 times ($3\times3\times3$) in each space direction and assuming full occupancy of the hydrate, i.e., there is a C$_2$H$_6$ molecule in each hydrate cage. As a result, the initial hydrate phase contains $1242$ molecules of water and 216 molecules of C$_2$H$_6$. Then, a pure water phase with 1242 molecules and a pure C$_2$H$_6$ guest phase with 400 molecules are added to the simulation box. Based on this setup, the initial configuration consists of a liquid water slab positioned between a solid slab of C$_2$H$_6$ hydrate on one side and a slab of liquid C$_2$H$_6$ molecules on the other. This arrangement guarantees the coexistence of all three phases under periodic boundary conditions. A schematic of the initial simulation box is presented in Fig.~\ref{snapshot}.

In all cases, the simulations are run using the $NPT$ or isothermal-isobaric ensemble, allowing each side of the simulation box to change independently to keep the pressure constant as well as to avoid any stress from the hydrate solid structure. We use the Verlet-leapfrog\cite{Cuendet2007a} algorithm for solving Newton's equations of motion with a time step of $2\,\text{fs}$. In order to keep the temperature and the pressure constant along the simulation, the v-rescale thermostat\cite{Bussi2007a} and the anisotropic Parrinello-Rahman barostat\cite{Parrinello1981a} are used with a time constant of $2\,\text{ps}$. In the case of the Parrinello-Rahman barostat, a compressibility value of $4.5\times10^{-5}\,\text{bar}^{-1}$ is applied in the three directions of the simulation box.

\section{Results}

We have performed simulations at seven different pressures, from $1000$ to $4000\,\text{bar}$, to determine the hydrate-water-C$_2$H$_6$ three-phase or dissociation line of the C$_2$H$_6$ hydrate. According to the DC technique, we simulate a set of temperatures for each pressure to locate the $T_{3}$ of the system. Particularly, for each pressure we examine temperature values near the experimental $T_3$, at a minimum interval of $2\,\text{K}$. This provides a margin of error for $T_3$ of at least $1\,\text{K}$ according to the exposed criterion. As we have already mentioned, each $T_3$ is estimated as the arithmetic average of the lowest temperature considered at which the ethane hydrate melts and the highest value at which the system freezes, with an uncertainty of $1\,\text{K}$. In all cases, we use the same initial simulation box generated as explained in Section II.

\begin{figure}
\includegraphics[width=\columnwidth]{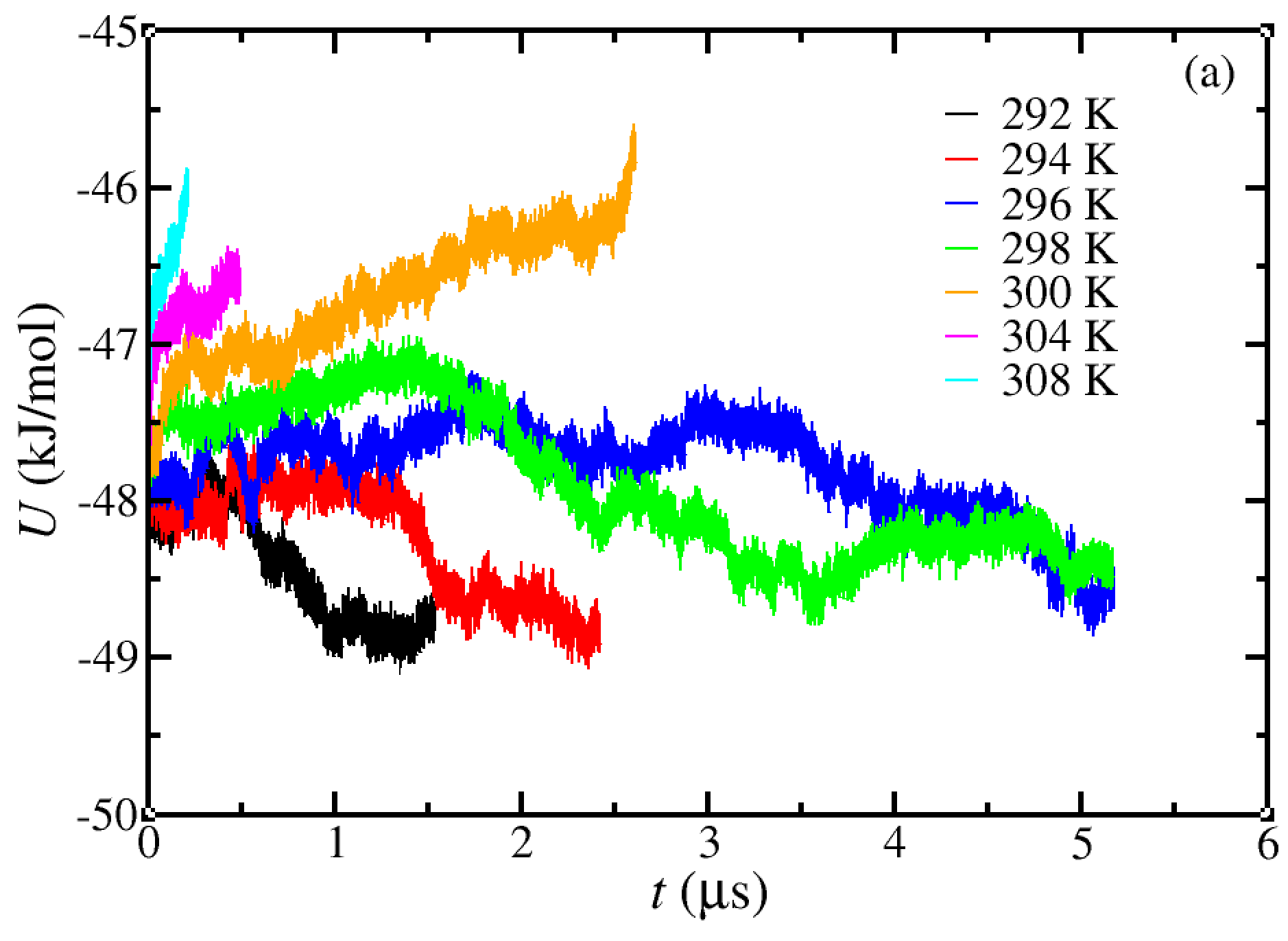} 
\includegraphics[width=\columnwidth]{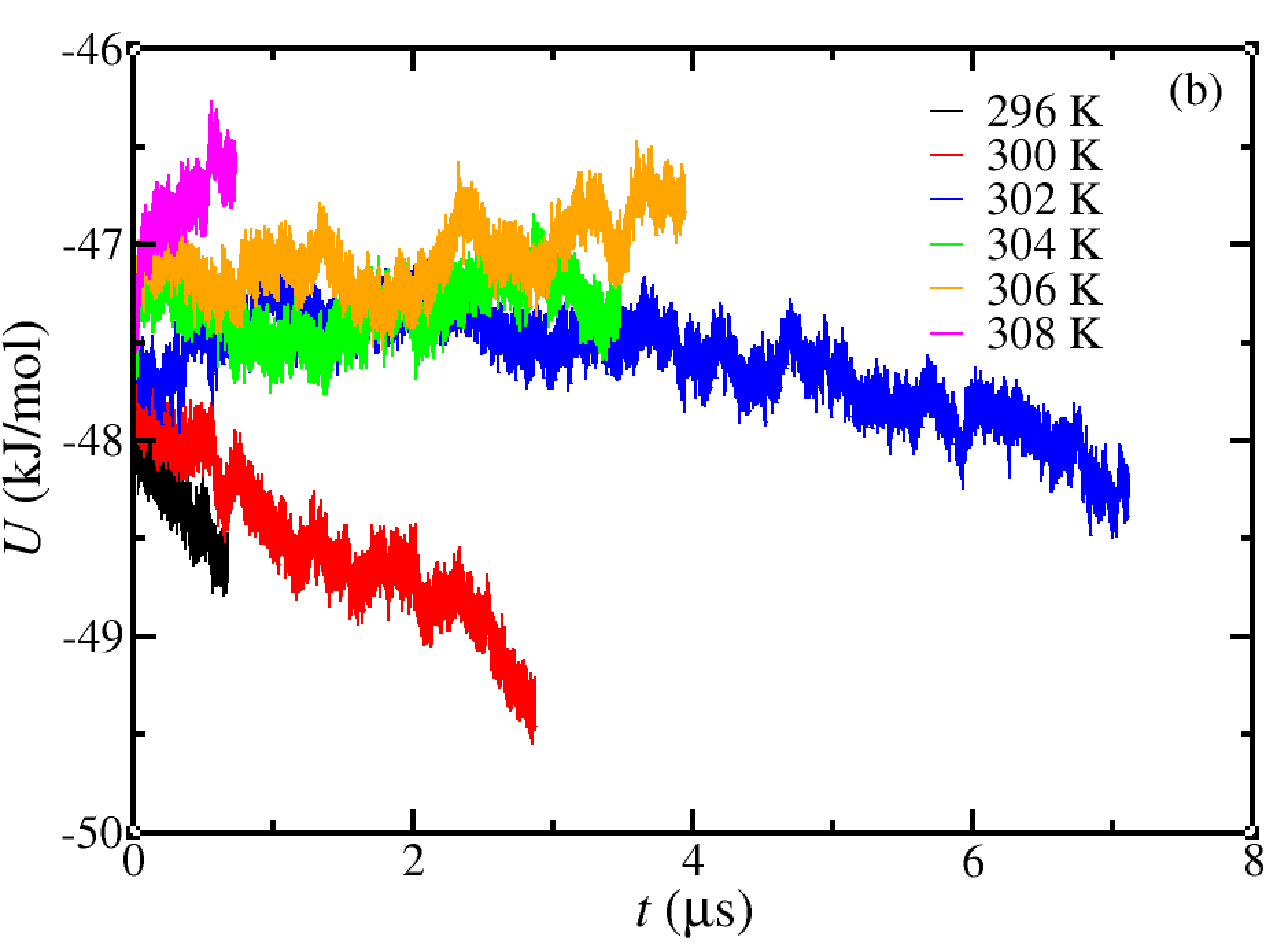} 
\includegraphics[width=\columnwidth]{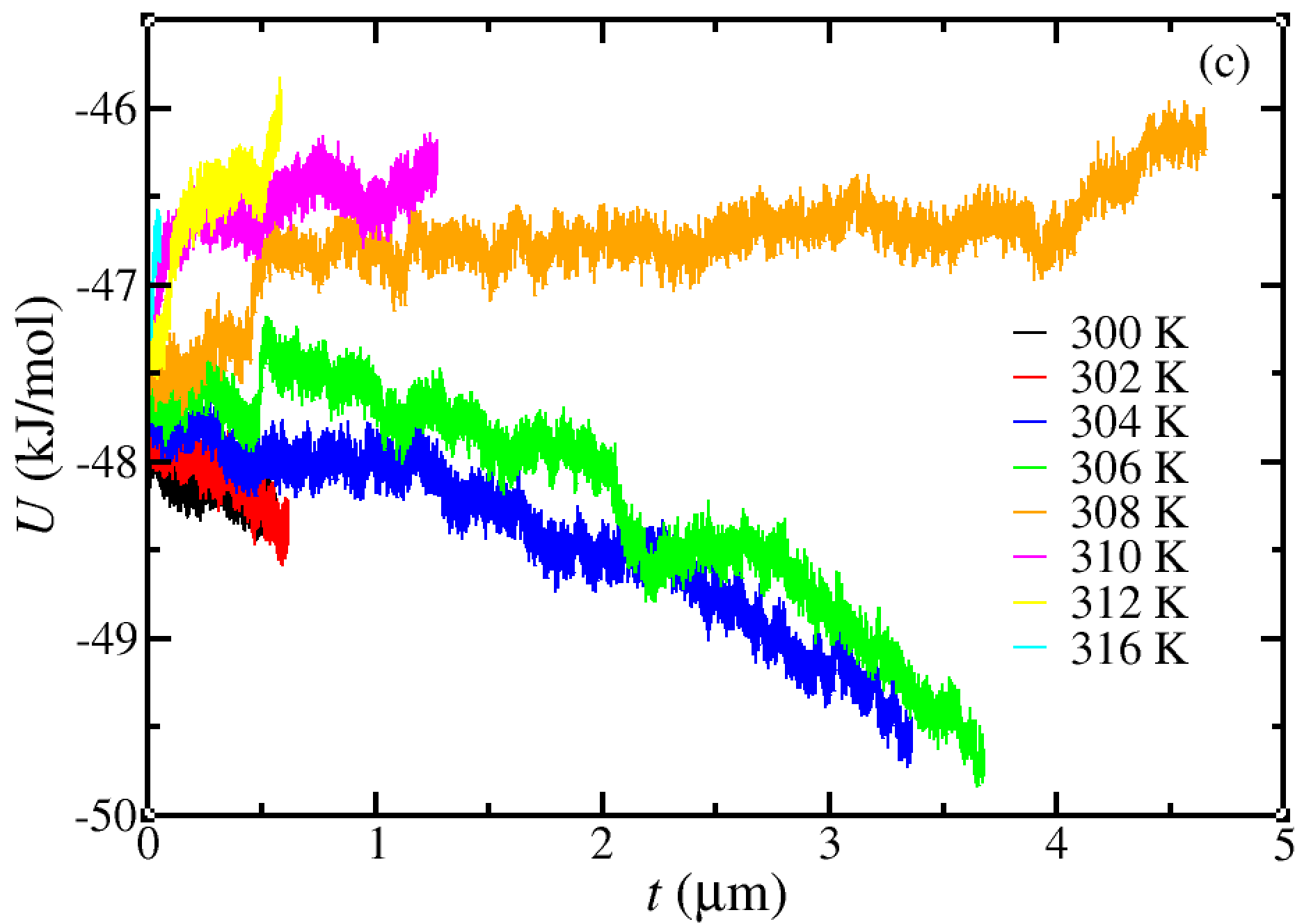} 
\caption{Evolution of the potential energy as a function of time for the \textit{NPT} runs of the three-phase system at 1000 bar (a),1500 bar (b), and 2000 bar (c) and various temperatures (see legends).}
\label{figure2}
\end{figure}

Figure~\ref{figure2} presents the time evolution of the system’s potential energy, $U$, at the lowest pressures examined in this study: $1000$, $1500$, and $2000\,\text{bar}$. We begin by analyzing the behavior at $1000\,\text{bar}$, shown in Fig.~\ref{figure2}a. At this pressure, simulations were conducted at seven temperatures ranging from $292$ to $308\,\text{K}$. As observed, the potential energy increases with time at temperatures above $300\,\text{K}$, signaling the dissociation of the C$_{2}$H$_{6}$ hydrate phase. In contrast, for temperatures below $298\,\text{K}$, the potential energy decreases, indicating crystallization of the hydrate. Based on these trends, the three-phase coexistence temperature, $T_{3}$, is estimated to lie between $298$ and $300\,\text{K}$. Applying this criterion, the dissociation temperature of the C$_{2}$H$_{6}$ hydrate at $1000\,\text{bar}$ is determined to be $299(1)\,\text{K}$, which is in excellent agreement with the experimental value reported in the literature at this pressure, $299.15\,\text{K}$.~\cite{Sloan2008a} We have also include these results in Table~\ref{table}.

Three important technical considerations must be addressed. First, although a rigorous statistical approach would require running multiple independent trajectories at each temperature, the computational cost of doing so is prohibitive for the purposes of this study. Given the extensive sampling needed across temperatures and pressures, we opted for single-trajectory simulations at each state point, which still provide valuable qualitative and quantitative insights into phase behavior. 

Second, the total simulation time varies depending on the target temperature. As the system approaches the three-phase coexistence temperature, $T_3$, its dynamics become increasingly stochastic, requiring longer trajectories to capture meaningful trends in the potential energy. For example, at higher temperatures such as $304$ and $308\,\text{K}$, the potential energy increases markedly within the first few hundred nanoseconds, consistent with hydrate dissociation. In contrast, for intermediate temperatures such as $296$ and $298\,\text{K}$, microsecond-scale simulations are necessary to fully observe the system’s evolution: initial trends may suggest hydrate melting, but extended simulations often reveal a reversal in potential energy, indicating hydrate growth. Depending on the thermodynamic conditions, the total simulation time ranges from hundreds of nanoseconds to several microseconds.

\begin{figure*}
\includegraphics[width=0.48\textwidth]{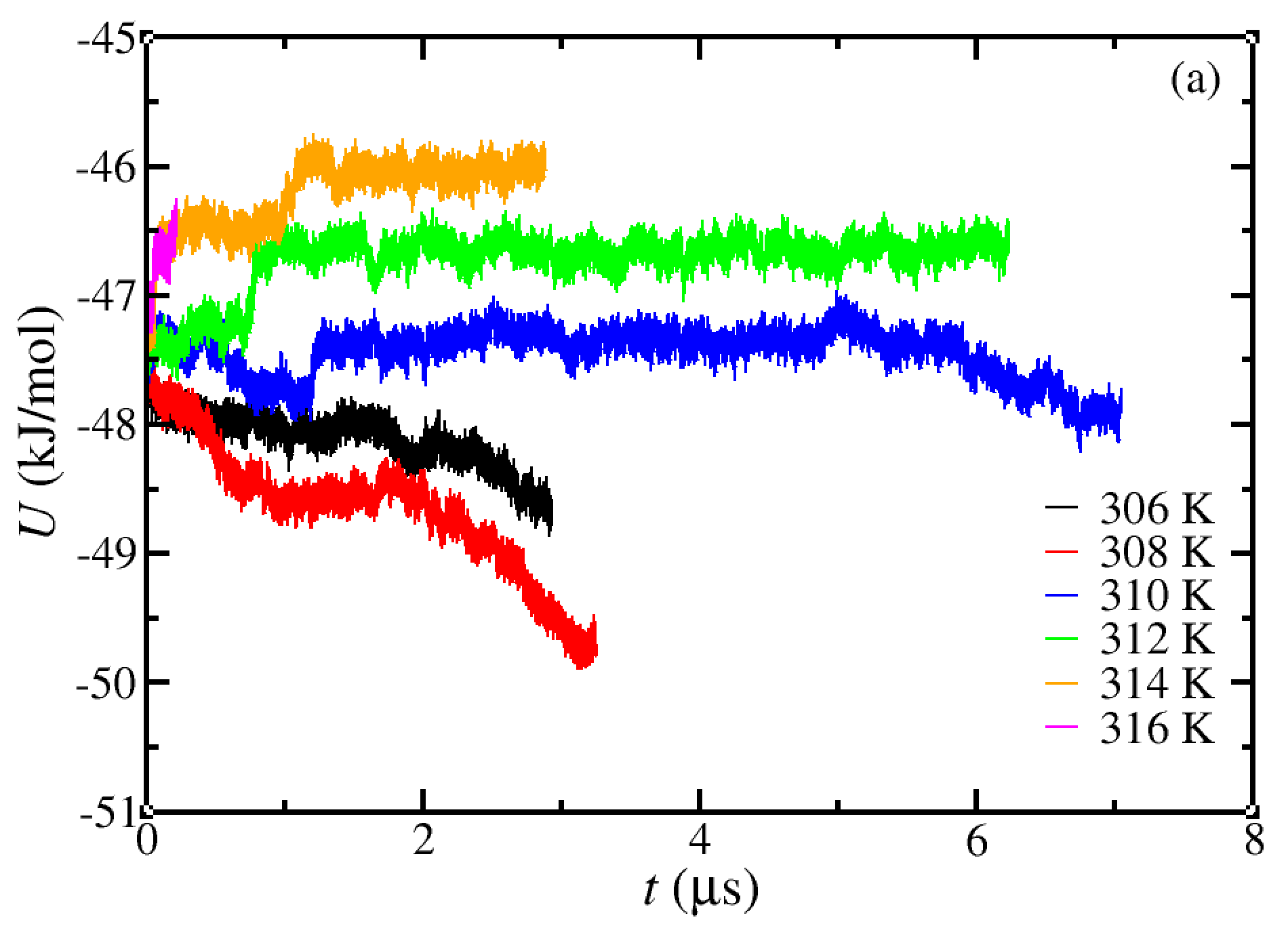} 
\includegraphics[width=0.48\textwidth]{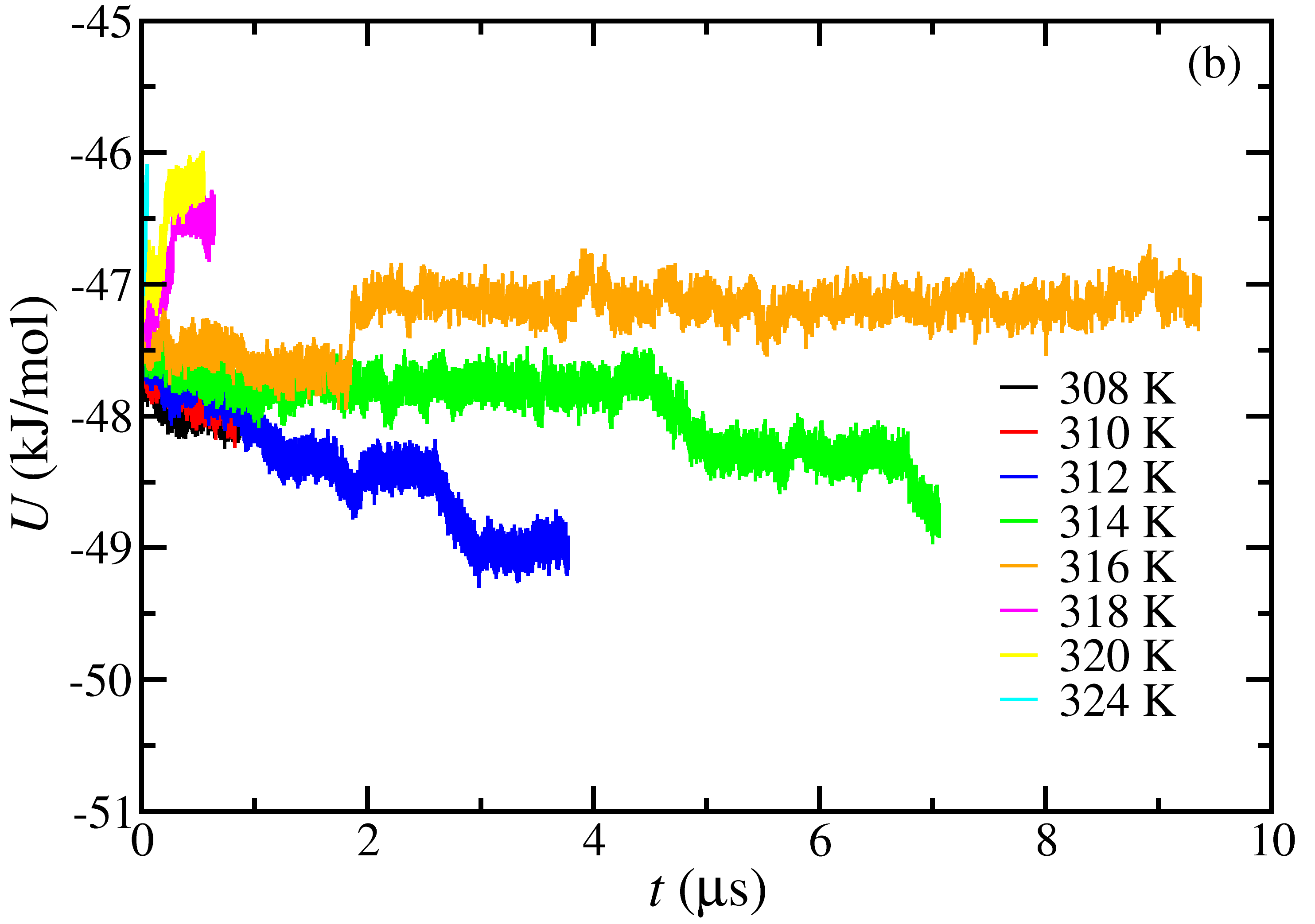} 
\includegraphics[width=0.48\textwidth]{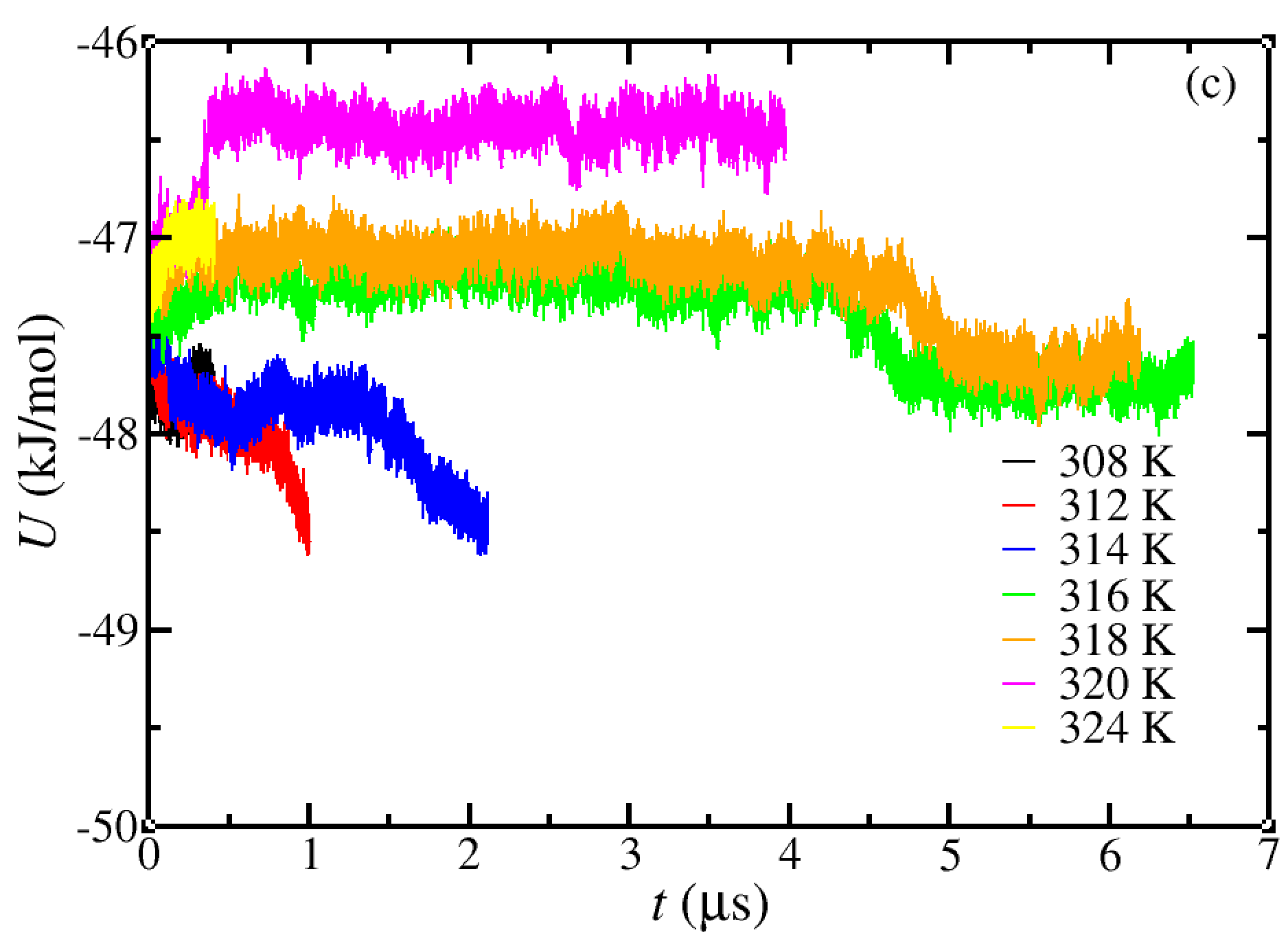} 
\includegraphics[width=0.48\textwidth]{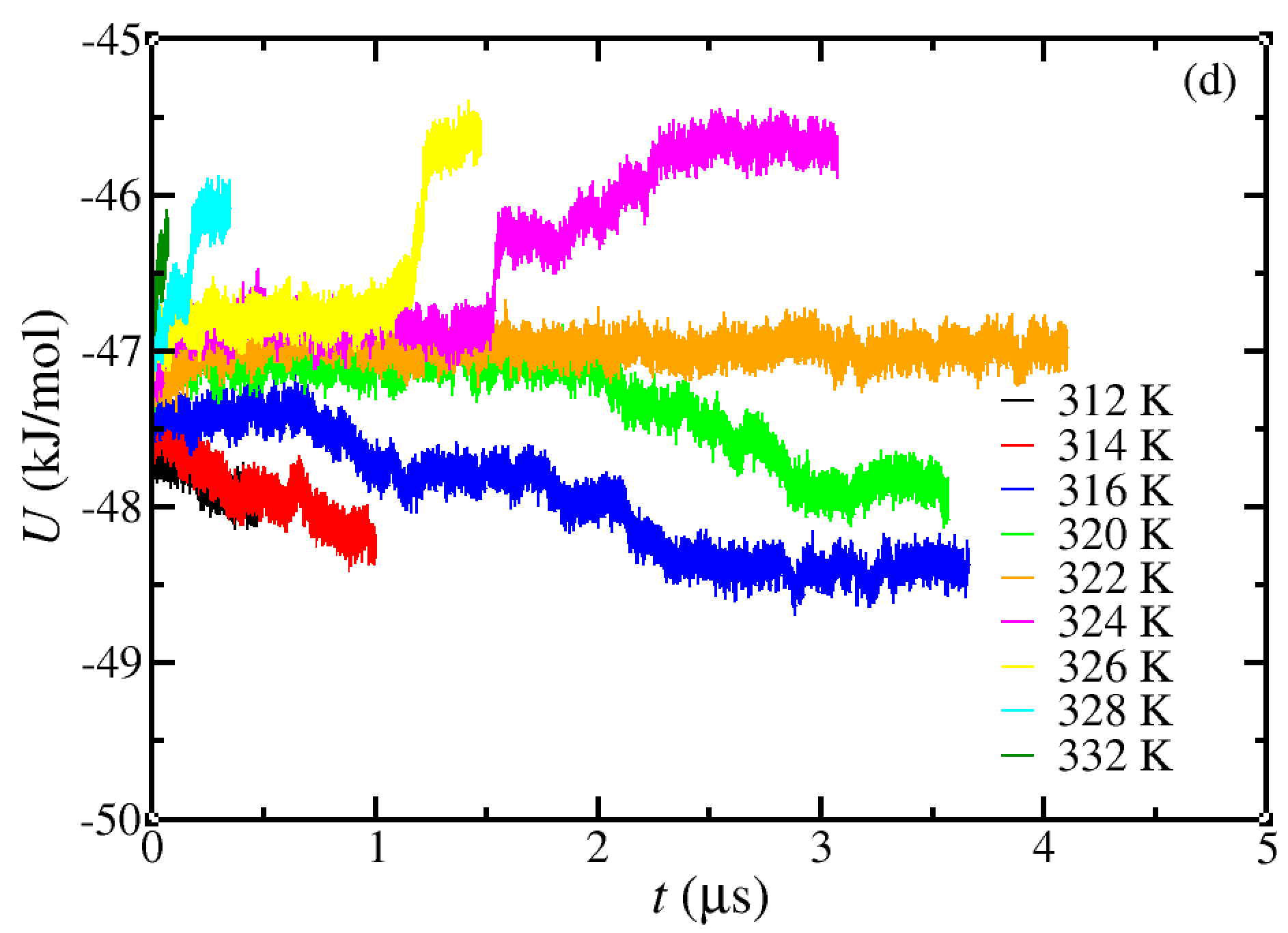} 
\caption{Evolution of the potential energy as a function of time as obtained from the \textit{NPT} runs of the three-phase system at 2500 (a), 3000 (b), 3500 (c), and $4000\,\text{bar}$ (d) and various temperatures (see legends).}
\label{figure3}
\end{figure*}

Third, the accuracy of the dissociation temperature predictions is supported by the use of the TIP4P/Ice water model,\cite{Abascal2005b} which is specifically parameterized to reproduce the experimental melting point of ice $I_{h}$. This choice is critical, as demonstrated in previous studies on CH$_{4}$ hydrates,\cite{Conde2013a} where it was shown that reliable determination of $T_3$ requires water models that accurately capture the thermodynamics of ice. In the present case, this model contributes significantly to the robustness of our predictions for ethane hydrate.

We now examine the results obtained at the second pressure investigated, $1500\operatorname{bar}$, as shown in Fig.~\ref{figure2}b. As in the previous case, the system exhibits a decrease in potential energy over time at $296$ and $300\operatorname{K}$, indicating crystallization of the hydrate phase. In contrast, at the highest temperature considered, $308\operatorname{K}$, the potential energy increases sharply, consistent with hydrate dissociation. For the intermediate temperatures between $302$ and $306\operatorname{K}$, longer simulation times are required to clearly resolve the system’s behavior. Based on the observed trends, the three-phase coexistence temperature, $T_3$, is estimated to lie between $304$ and $306\operatorname{K}$. From this range, we determine $T_3=305(1)\operatorname{K}$, which is in excellent agreement with the experimental value reported at this pressure, $304.15\operatorname{K}$.\cite{Sloan2008a} (see also Table~\ref{table}).

Following the same approach used for the two previous pressures, we analyze the time evolution of the system’s potential energy, $U$, at $2000\operatorname{bar}$, as shown in Fig.~\ref{figure2}c. Simulations are performed at eight different temperatures. As clearly observed, the C$_{2}$H$_{6}$ hydrate phase fully dissociates at temperatures above $308\operatorname{K}$, while for temperatures below $306\operatorname{K}$, the potential energy decreases over time, indicating the formation of the hydrate phase from liquid water and ethane. Based on these observations, the three-phase coexistence temperature, T$_{3}$, is estimated to be
$307(1)\operatorname{K}$, in excellent agreement with the experimental value of $307.26\operatorname{K}$.~\cite{Sloan2008a} This results is also included in Table~\ref{table}. Notably, at this pressure, the system’s energy evolution-whether increasing or decreasing—is more readily distinguishable from the early stages of the simulations, in contrast to the longer onset times required at lower pressures.

\begin{figure*}
\includegraphics[valign=m,width=0.48\textwidth]{figure4a.eps}
\includegraphics[valign=m,width=0.48\textwidth]{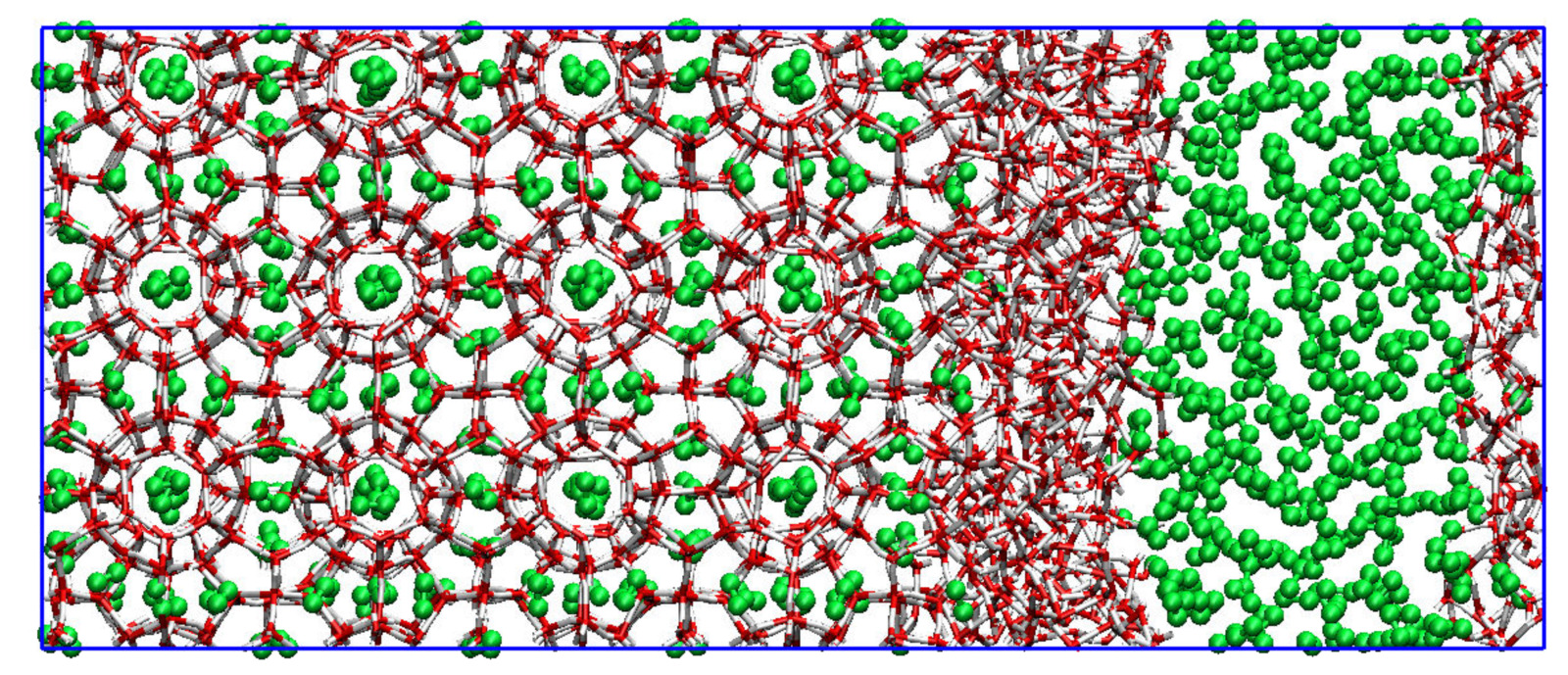}
\includegraphics[valign=m,width=0.48\textwidth]{figure4c.eps}
\includegraphics[valign=m,width=0.48\textwidth]{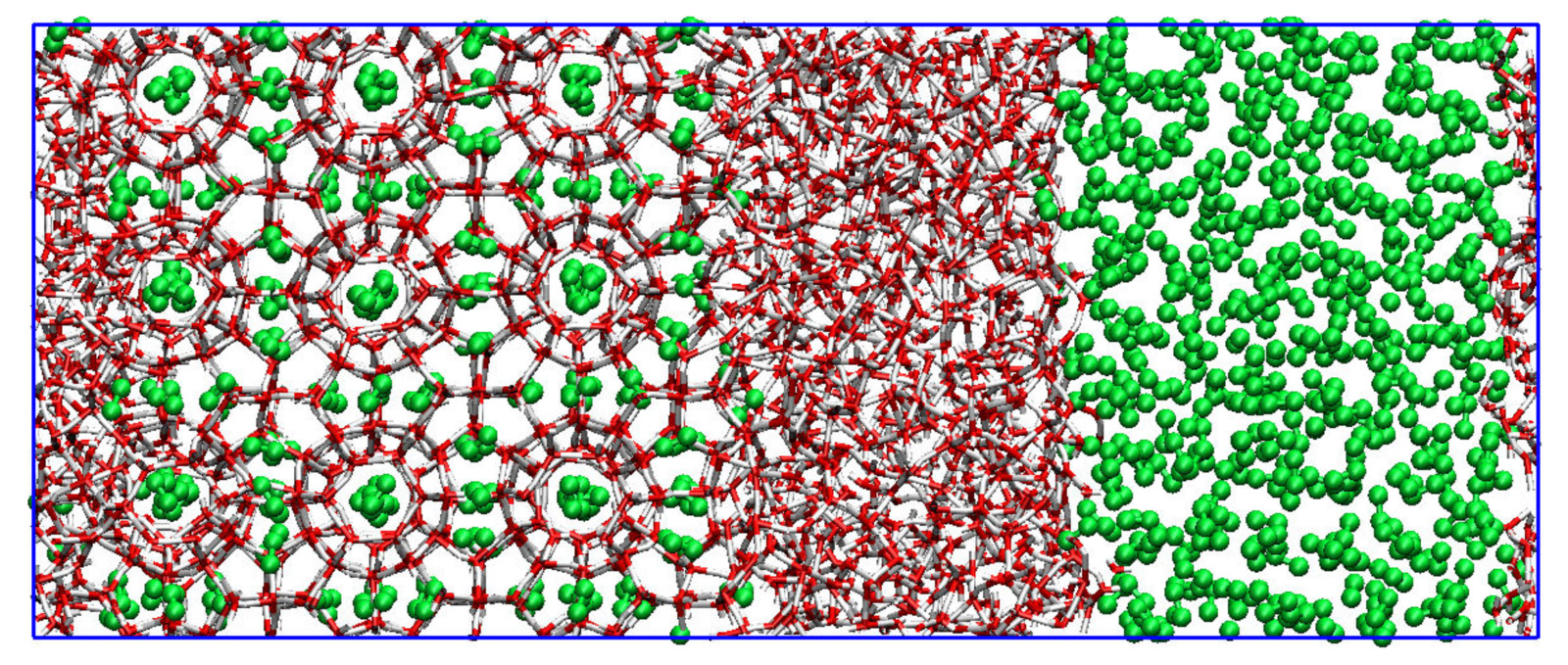}
\includegraphics[valign=m,width=0.48\textwidth]{figure4e.eps}
\includegraphics[valign=m,width=0.48\textwidth]{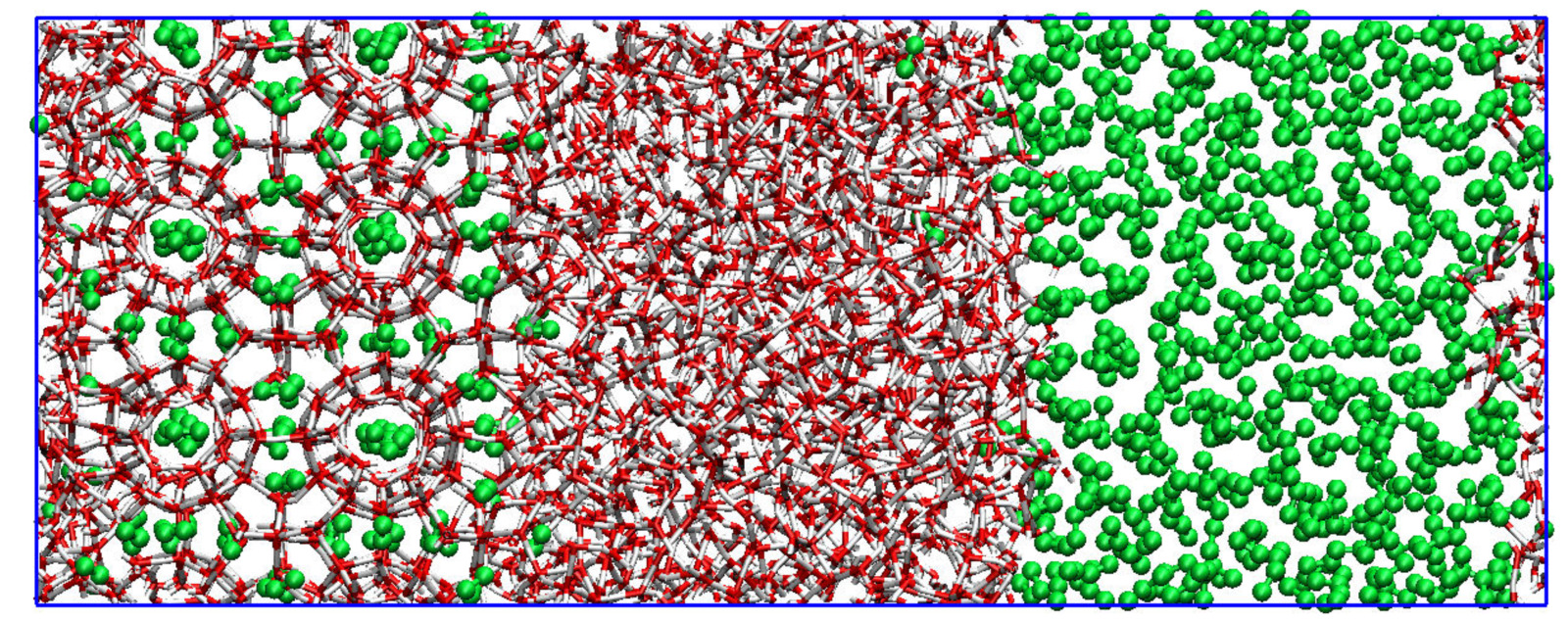}
    \caption{Ethane and water density profiles and snapshots at $2500\operatorname{bar}$ at 308 (top), 310 (middle), and $312\operatorname{K}$ (bottom). The density profiles show the initial and final density distribution of water and C$_2$H$_6$ along the simulation box, while the snapshots show the final configuration obtained from the $NPT$ simulations.}
    \label{figure4}
\end{figure*}

Finally, we examine the dissociation behavior of the C$_{2}$H$_{6}$ hydrate at the highest pressures considered in this study: $2500$, $3000$, $3500$, and $4000\operatorname{bar}$. The time evolution of the system’s potential energy at various temperatures for each pressure is shown in Fig.~\ref{figure3}. As with the lower pressures, a consistent trend is observed: at higher temperatures, the hydrate dissociates, while at lower temperatures, it forms. The dissociation temperature at each pressure lies between these two regimes. Specifically, at $2500\operatorname{bar}$, dissociation occurs between $310$ (blue) and $312\operatorname{K}$ (green); at $3000\operatorname{bar}$, between $314$ (green) and $318\operatorname{K}$ (orange); at $3500\operatorname{bar}$, between $318$ (orange) and $320\operatorname{K}$ (magenta); and at $4000\operatorname{bar}$ between $320$ (green) and $324\operatorname{K}$ (magenta). From these observations, the estimated three-phase coexistence temperatures are $311(1)$, $316(2)$, $319(1)$, and $322(2)\operatorname{K}$ for $2500$, $3000$, $3500$, and $4000\operatorname{bar}$, respectively. All these results have also been included in Table~\ref{table}.

It is worth noting that the estimated uncertainty in the simulated dissociation temperatures is $1\operatorname{K}$ for all pressures except for $3000$ and $4000\operatorname{bar}$, where it increases to $2\operatorname{K}$. At $3000\operatorname{bar}$ (Fig.~\ref{figure3}b), the three-phase equilibrium temperature lies between $314\operatorname{K}$—where the potential energy decreases (green curve)—and $318\operatorname{K}$, where it increases (magenta curve). Although simulations were extended to nearly $10\mu\operatorname{s}$, the evolution of the potential energy at $316\operatorname{K}$ (orange) remained inconclusive, and neither the time series nor the corresponding density profiles provided clear evidence of phase behavior. Consequently, the dissociation temperature at this pressure is estimated to lie within the range of $314-318\operatorname{K}$, yielding $T_3=316(2)\operatorname{K}$.

At $4000\operatorname{bar}$ (Fig.~\ref{figure3}c), potential energy decreases are observed from the beginning of the simulations at $312$ and $314\operatorname{K}$, suggesting hydrate formation. However, for temperatures between $316$ and $320\operatorname{K}$, longer simulations were required. Beyond $2000\operatorname{ns}$, the potential energy decreases for both $316$ and $320\operatorname{K}$, indicating crystallization, whereas at $322\operatorname{K}$, the potential energy remains nearly constant and the density profiles do not clearly indicate either melting or growth. Based on these observations, the dissociation temperature at $4000\operatorname{bar}$ is estimated as 
$T_3=322(2)\operatorname{K}$.

\begin{table}
\caption{Three-phase dissociation temperatures, $T_3$, of C$_{2}$H$_{6}$ hydrate obtained from molecular simulations at various pressures, $P$. The last column reports the corresponding experimental values, $T^{exp}$, from the literature.~\cite{Sloan2008a}}
 \centering
\begin{tabular}{lccccc}
\hline\hline
$P$ (bar) & $T$ (K) & $T^{exp}$ (K)\\
\hline
1000  & 299 (1) & 299.15\\
1500  & 305 (1) & 304.15\\
2000  & 307 (1) & 307.26\\
2500  & 311 (1) & 311.40\\
3000  & 316 (2) & 314.20\\
3500  & 319 (1) & 317.49\\
4000  & 322 (2) & 319.65\\
\hline\hline
\end{tabular}
\label{table}
\end{table}

Having presented the results at each pressure, we now turn to a general analysis of the time evolution of the system’s potential energy and its relation to hydrate formation and dissociation across the studied pressure range. Overall, the slope of the decreasing potential energy curves indicates that hydrate growth becomes progressively slower with increasing pressure, suggesting a pressure-dependent kinetic barrier to crystallization. In contrast, hydrate melting tends to occur more rapidly at higher pressures, as evidenced by the earlier onset of energy increases. Additionally, the simulated three-phase coexistence temperatures, $T_{3}$, show excellent agreement with experimental data at low and intermediate pressures, while a slight overestimation is observed at the highest pressures investigated.

As previously noted, some of the simulated systems—particularly those at $2500$, $3000$, and $3500\operatorname{bar}$—exhibit significant challenges in reaching equilibrium, complicating the identification of the final state (melting or freezing). To gain deeper insight into the system's evolution under these conditions, we focus on representative simulations of C$_2$H$_6$ hydrates at $2500\operatorname{bar}$ (Fig.~\ref{figure3}b). Note that the same procedure has been used for the rest of the pressures in this work. We concentrate on the key temperatures of $308$ (red), $310$ (blue), and $312\operatorname{K}$ (green). In order to unambiguously determine whether the system tends toward melting or freezing, we analyze the density profiles of both C$_2$H$_6$ and water across the three-phase regions at these selected temperatures, as shown in Fig.~\ref{figure4}. To complement this analysis, we also include representative snapshots for each configuration, which illustrate the final state of the system and facilitate a visual comparison. For context, it is helpful to compare these snapshots in Fig.~\ref{figure4} with the initial three-phase configuration shown in Fig.~\ref{snapshot}.

As shown in Fig.~\ref{figure4}, the system exhibits both growth and melting of the hydrate layer, consistent with the trends previously inferred from the potential energy versus time curves. At $308\operatorname{K}$ (top density profiles and corresponding snapshot), the fluid phases undergo freezing, leading to a clear expansion of the initial ethane hydrate slab. In particular, the number of water and ethane layers in the hydrate phase increases significantly from the initial configuration (black and green curves) to the final stage of the simulation (red and blue curves). Initially, the hydrate–water interface is located at approximately $3.5\operatorname{nm}$, but after $3.2\mu\text{s}$, it has advanced to around $6\operatorname{nm}$. Concurrently, the water-rich phase, initially spanning from $3.5$ to $5.5\operatorname{nm}$, has contracted to a width of about $1\operatorname{nm}$ or less. The ethane-rich liquid phase also exhibits a slight reduction in size, although to a lesser extent, as reflected in the density profiles. This behavior is further corroborated by the visual comparison of the final system snapshot at $312\operatorname{K}$ (top panel in Fig.~\ref{figure4}) with the initial three-phase configuration (Fig.~\ref{snapshot}). The hydrate phase has clearly expanded—with the formation of an additional layer composed of large or T cages and some small or D cages—while the liquid water and ethane phases have notably diminished, especially the water phase.

At $312\operatorname{K}$, the system exhibits a markedly different behavior, as shown in the lower density profiles and corresponding snapshot. The hydrate phase undergoes dissociation, resulting in a noticeable reduction of the initial hydrate slab. This is evidenced by the decrease in the number of water and ethane layers from the initial configuration (black and green curves) to the final state (red and blue curves). Over the course of the $6.2\mu\text{s}$ simulation, the hydrate–water interface shifts from approximately $3.5$ to $2.5\operatorname{nm}$, indicating the retreat of the solid phase. Simultaneously, the water-rich region expands from an initial width of $2$ to about $3\operatorname{nm}$, while the ethane-rich liquid phase also grows slightly. These structural changes are further supported by the final system snapshot (bottom panel, Fig.~\ref{figure4}), which shows the contraction of the hydrate and the corresponding expansion of the surrounding fluid phases. Notably, a layer composed of large (T) cages and some small (D) cages disappears, highlighting the breakdown of the crystalline structure.

The middle density profiles and corresponding snapshot in Fig.~\ref{figure4} depict the final configuration obtained at $312\operatorname{K}$. As indicated by the potential energy evolution in Fig.~\ref{figure3}a (blue curve), the system exhibits a net decrease in energy over time, suggesting hydrate growth. Although the structural changes are less pronounced than those observed at $308\operatorname{K}$, both the density profiles and the snapshot provide evidence of hydrate formation. Specifically, the water density profile (red curves) shows increased structuring, and the snapshot reveals the early development of a hydrate layer, originating from a thin layer of small (D) cages at the hydrate–water interface. Considering that the hydrate melts at $312$ and grows at $310\operatorname{K}$, the three-phase coexistence temperature is estimated as $311(1)\operatorname{K}$, in excellent agreement with the experimental value of $311.4\operatorname{K}$.\cite{Sloan2008a} This conclusion is consistent with the interpretation of the potential energy curves in Fig.~\ref{figure3}b.

\begin{figure}
\includegraphics[width=\columnwidth]{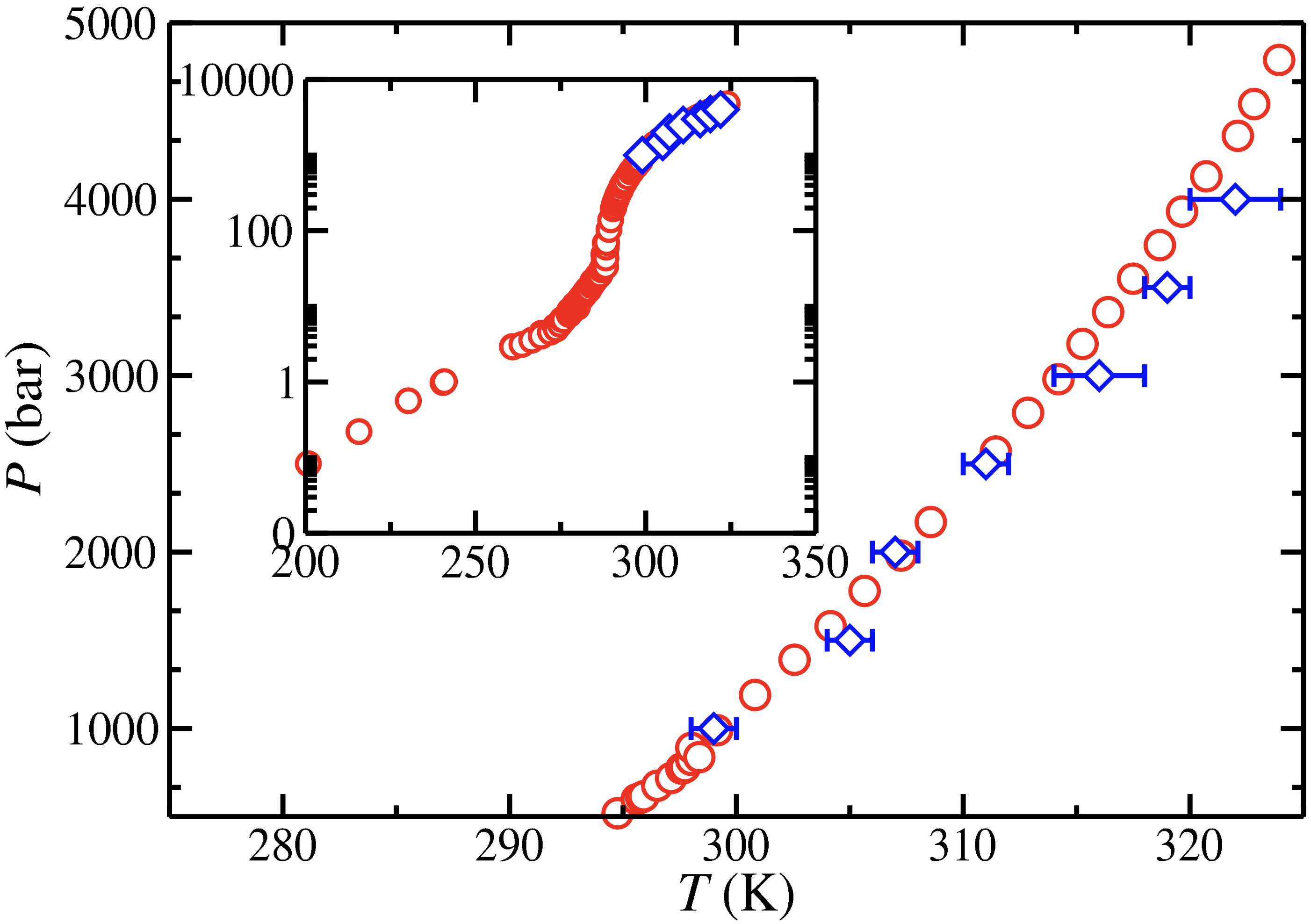}   
\caption{Pressure-temperature projection of the dissociation line of the C$_{2}$H$_6$ hydrate. Blue diamonds are the results obtained in this work using the direct coexistence method, the TIP4P/Ice model for water, and the TraPPE model for C$_{2}$H$_6$. Red circles correspond to experimental data taken from the literature.~\cite{Roberts1940a,Deaton1946a,Reamer1952a,Galloway1970a,Falabella1974a,Holder1980a,Holder1982a,Ng1985a,Avlonitis1988a,Song1989a,Nakano1998b,Yang2000a,Morita2000a}}
\label{figure5}
\end{figure}


As it has been commented previously, Morita and co-workers demonstrated that C$_2$H$_6$ molecules can occupy both types of cages~\cite{Morita2000a} although it had long been assumed that C$_{2}$H$_{6}$ was too large to fit into the smaller D cages, and only the big T cages are filled.~\cite{Sloan2008a} In this work, and following the work of Morita \emph{et al.},~\cite{Morita2000a} the initial hydrate seed is fully occupied. However, from the analysis of the density profiles of those simulations where the hydrate phase grows, it is possible to analyze the occupancy of the new-growth hydrate. The C$_2$H$_6$/water molecule ratio in a fully occupied hydrate is 8/46, where 6 C$_2$H$_6$ molecules occupy the large T cages and 2 C$_2$H$_6$ molecules occupy the small D cages. If only the large T cages of the new-growth hydrate are occupied by C$_2$H$_6$ molecules, while the small D cages remain empty, the C$_2$H$_6$/water molecule ratio would be 6/46. In this work, we have analyzed the density at one temperature at the lowest pressure ($1000\,\text{bar}$), an intermediate pressure ($2500\,\text{bar}$), and the highest pressure ($4000\,\text{bar}$). From these pressures, we have selected the density profile with the highest expansion of the hydrate phase from the different simulated temperatures. This is done in order to ensure that the new-growth hydrate phase is large enough to provide enough statistics about the number of C$_2$H$_6$ and water molecules. In particular, we have analyzed the density profiles at $1000\,\text{bar}$ and $298\,\text{K}$, $2500\,\text{bar}$ and $308\,\text{K}$, and $4000\,\text{bar}$ and $320\,\text{K}$. The C$_2$H$_6$/water molecule ratios of the new-growth hydrate obtained at these conditions are 7.8/46, 8.2/46, and 7.9/46. Notice that in all cases, we can round the C$_2$H$_6$/water molecule ratio to 8/46, which corresponds to a fully occupied sI hydrate as Morita \emph{et al.}~\cite{Morita2000a} claimed in their experimental work. However, these calculations have to be considered just as a first approximation to understand the effect of the occupancy of the small D cages, and further work is required to understand it fully. A more rigorous determination of the hydration number of hydrates is possible,~\cite{Handa1986a,Qin2013a,Udachin2002a,Takeya2010a} but this lies beyond the scope of the present work.

Finally, the dissociation line of C$_{2}$H$_{6}$ hydrate obtained from molecular simulations is summarized in the pressure–temperature phase diagram shown in Fig.~\ref{figure5}. For comparison, experimental data from the literature covering the entire pressure range studied are also included.~\cite{Roberts1940a,Deaton1946a,Reamer1952a,Galloway1970a,Falabella1974a,Holder1980a,Holder1982a,Ng1985a,Avlonitis1988a,Song1989a,Nakano1998b,Yang2000a,Morita2000a} It is interesting to note that the C$_2$H$_6$-water phase diagram shows two quadrupole points.~\cite{Sloan2008a} The first (lower) quadrupole point, Q$_1$, occurs at $272.9\,\text{K}$ and $25.6\,\text{bar}$. Under these conditions, there exist four phases in equilibrium: a hydrate phase, an ice Ih phase, a water-rich liquid phase, and a vapor phase. The second (upper) quadrupole point, Q$_2$, occurs at $287.8\,\text{K}$ and $33.9\,\text{bar}$. Under these conditions, there exist four phases in equilibrium: a hydrate phase, a water-rich liquid phase, a C$_2$H$_6$-rich liquid phase, and a vapor phase. The changes in slope occur precisely under these thermodynamic conditions, corresponding to the ice Ih–to–water-rich phase transition and the vapor–liquid transition of the C$_2$H$_6$-rich phase. As observed, the simulated dissociation line agrees well with the experimental results, with all values falling within the estimated statistical uncertainties reported throughout the text. Notably, both the simulated and experimental curves exhibit similar slopes, indicating that the agreement is not only quantitative but also qualitative.

\section{Conclusions}

In this work, we have carried out molecular dynamics simulations employing the direct coexistence method to determine the three-phase (hydrate–water–ethane) coexistence line of the C$_{2}$H$_{6}$ hydrate system. Simulations are performed at pressures ranging from $1000$ to $4000\operatorname{bar}$. The predicted dissociation temperatures, $T_{3}$, show very good agreement with experimental data across the entire pressure range, with only slight overestimations observed at the highest pressures. These results highlight the reliability of the employed methodology and interaction models in capturing the phase behavior of this computationally underexplored system. Specifically, the combination of the TIP4P/Ice water model with the TraPPE force field for ethane, using Lorentz–Berthelot combining rules for cross interactions, proves to be a robust and accurate choice. No adjustments to the standard combining rules or inclusion of long-range Lennard-Jones corrections are necessary to achieve quantitative agreement. Additionally, we find that the simulation time required to observe phase transitions strongly depends on the proximity to the dissociation temperature: significantly longer trajectories are needed when the system is near $T_{3}$, due to the increasingly slow dynamics near equilibrium.

\section*{Acknowledgements}
The authors wish to dedicate this work and express their heartfelt gratitude to our friend Prof.~Carlos Vega for his invaluable contributions to the study of complex systems by computer simulation, and in particular for his efforts to understand the properties of water from a molecular perspective, including hydrate clathrates. The development of simplified yet accurate water models over the past decades has been fundamental to advancing microscopic knowledge of aqueous systems, providing deeper insight into their structure and behavior. His legacy continues to guide both theoretical and applied research, inspiring future generations of scientists not only in Spain but throughout the world. We deeply value his tireless dedication and the enduring impact of his contributions to science. In the group at the University of Huelva, we are deeply grateful for your sincere friendship, which goes beyond scientific cordiality and reaches a personal level. Carlos, you know that you are always welcome in Huelva by all of us and by the V. de R. We acknowledge Grant Refs (PID2021-125081NB-I00 and PID2024-158030NB-I00) financed both by MCIN/
AEI/10.13039/501100011033 and FEDER EU, and Universidad de Huelva (P.O. FEDER EPIT1282023), also cofinanced
by EU FEDER funds. M.J.T. acknowledges the research contract (Ref 01/2022/38143) of Programa
Investigo (Plan de Recuperación, Transformación y Resiliencia, Fondos NextGeneration EU) from Junta de Andalucía (HU/INV/0004/2022). Part of the computations was carried out at the Centro de Supercomputación de Galicia (CESGA, www.cesga.es, Finisterrae III Supercomputer). We also greatly acknowledge RES resources provided by the Barcelona Supercomputing Center in Mare Nostrum to FI-2025-2-0032.

\section*{Author declarations}

\noindent
\textbf{Conflict of interests}

The authors declare no conflicts to disclose.

\section*{Author contributions}

\noindent
\textbf{Paula Gómez-Álvarez:} Methodology (equal); Investigation (lead); Writing – original draft (equal); Writing – review \& editing (equal).\textbf{Miguel J. Torrejón:} Methodology (equal); Writing – original draft (equal); Writing – review \& editing (equal). \textbf{Jesús Algaba:} Methodology (equal); Writing – original draft (equal); Writing – review \& editing (equal).
\textbf{Felipe J. Blas:} Conceptualization (lead); Funding acquisition (lead); Methodology (equal); Writing – original draft (equal); Writing – review \& editing (equal).

\section*{Data availability}

The data that support the findings of this study are available within the article.

\bibliography{masterbib}

\end{document}